# Low-frequency interlayer phonon dynamics and photoinduced terahertz absorption in black phosphorus

Haiyun Huang[1,2,3], Cheng Chen[4], Yuxin Zhai[5], Xiu Zhang[1,2,3], Junzhi Zhu[1,2,3], Junyong Wang[4], Kai Zhang[4], Qihua Xiong[1,5,6,7,*], and Haiyun Liu[1,†]

[1] Beijing Academy of Quantum Information Sciences, Beijing 100193, P.R. China;

[2] Beijing National Laboratory for Condensed Matter Physics, Institute of Physics, Chinese Academy of Sciences, Beijing 100190, P. R. China;

[3] School of Physical Sciences, University of Chinese Academy of Sciences, Beijing 100049, P. R. China;

[4] Key Laboratory of Semiconductor Display Materials and Chips & i-Lab, Suzhou Institute of Nano-Tech and Nano-Bionics (SINANO), Chinese Academy of Sciences, Suzhou, Jiangsu 215123, China.

[5] State Key Laboratory of Low-Dimensional Quantum Physics and Department of Physics, Tsinghua University, Beijing 100084, P.R. China;

[6] Frontier Science Center for Quantum Information, Beijing 100084, P. R. China;

[7] Collaborative Innovation Center of Quantum Matter, Beijing 100084, P.R. China.

*Contact author: qihua_xiong@tsinghua.edu.cn

†Contact author: liuhy@baqis.ac.cn

**Abstract**

The strong interlayer coupling in black phosphorus (BP), arising from wavefunction overlap between layers, is critical for understanding its electronic and optical properties. Here, we utilize terahertz (THz) spectroscopy to study phonon dynamics in BP. We identify two peaks at 6 and 8.5 meV in steady-state THz spectra, which are attributed to low-frequency interlayer phonon modes. Both modes exhibit anharmonic phonon coupling behavior below 150 K, manifesting as temperature-dependent red-shifts. Using time-resolved THz spectroscopy, we observe significantly increased THz absorption under photoexcitation, arising from the transient enhancements of the extended Drude component and the oscillator strengths of interlayer phonons in non-equilibrium. These findings highlight the critical role of interlayer phonon dynamics in understanding many-body physics in BP.

## 1. Introduction

Black phosphorus (BP) is a layered elemental semiconductor that has been attracting significant attention due to its notable properties, including sizable bandgaps [1,2,3], excellent carrier mobility [4,5,6], high on-off current ratio [7], well-developed current saturation [8,9], and suitable integration in devices [10,11,12]. The direct bandgap is maintained from bulk to monolayer BP. The bandgap of bulk BP is approximately 0.3 eV. The reduction of layer number leads to the increase of the bandgap due to quantum confinement effects, ultimately reaching nearly 2 eV in the monolayer limit [2,13]. In contrast to gapless graphene and transition metal dichalcogenides (TMDs) with large bandgaps in the near-infrared or visible region, BP possesses layer-controlled bandgaps spanning from the mid-infrared to the visible spectrum. This unique electronic characteristic positions BP as a material that bridges the gap between graphene and TMDs [14]. In addition, the electronic and optical properties of BP exhibit significant sensitivity to several external factors, such as doping [15,16], orientation [17,18] and strain [17,19,20,21,22]. Moreover, the heterojunction interface engineering based on BP promises multifunctional optical and electronic properties [23,24]. Owing to these advantages, BP holds tremendous potential for the applications in electronic and optoelectronic technologies [25,26,27].

Under ambient conditions, the most stable crystal structure of BP adopts an orthorhombic symmetry, characterized by layers of phosphorus atoms arranged in a puckered honeycomb network [28]. As illustrated in Fig. 1(a), the puckered lattice of BP forms two inequivalent in-plane directions: the x-axis along the armchair direction and the y-axis along the zigzag direction. In addition to the weak van der Waals (vdW) interaction between atomic layers, which is a dominant feature in many 2D materials, an important recognition for BP lies in the strong interlayer coupling. Such a coupling is understood to induce the thickness-dependent lattice and band parameters [5,29]. The wavefunction overlap of the conduction band minimum and the valence band maximum under bias, resulting from the strong interlayer electronic state coupling, gives rise to significant dipole oscillator strength, enabling bandgap tuning properties [30]. Notably, the interlayer coupling in BP activates dark exciton [31], drives band renormalization [5,32], and introduces a pronounced temperature dependence of the band structure [13]. Therefore, investigating the interlayer coupling is of critical importance for understanding electronic and optical properties of BP.

Many-body effects, including electron-electron, electron-phonon and phonon-phonon interactions, play crucial roles in determining carrier and optical properties of materials, making them essential for practical applications [24,33,34]. Investigations of phonon dynamics and electron-phonon coupling in BP are fundamentally important, given that stress-induced lattice deformation can effectively alter the band structure and even induce a semiconductor-to-metal transition [19]. External pressure can induce Lifshitz transition [35,36] and even create a superconducting transition with a critical temperature higher than 10 K [37,38]. Theoretical calculations predict that phonon softening by the electron-doping leads to the strong electron-phonon coupling in monolayer BP, and induces superconductivity that can be enhanced by strain [39]. Further investigations of BP have revealed that phonons significantly influence the thermal conductivity [40] and the exciton properties [41]. Different from the traditional intralayer high-frequency phonon modes, the rigid-layer vibrations in BP generate low-frequency interlayer phonons, such as in-plane shear and out-of-plane breathing modes, with frequencies below 100 cm$^{-1}$ (12.4 meV) [29,42,43,44,45]. Hence, low-frequency interlayer phonons serve as a direct indicator of the interlayer coupling strength. However, systematic experimental studies of low-frequency interlayer phonons, particularly through terahertz (THz) spectroscopy, are still lacking.

THz spectroscopy is a powerful tool for studying charge carrier dynamics and low-frequency phonon resonances owing to the THz energy range covering from few to 100 meV [46]. It brings pivotal advantages in contrast to traditionally incoherent far-infrared measurements, because the THz time-domain spectroscopy (TDS) is coherently detected, enabling directly extraction of the complex optical conductivity without involving the Kramers-Kronig (KK) transformation [47]. Moreover, time-resolved THz spectroscopy, using an optical-pump THz-probe scheme, enables the time tracking of charge carrier excitation, cooling and recombination processes, as well as the dynamics of low-energy excitations [23,46,47,48].

In this study, we employ steady-state and time-resolved THz spectroscopy to investigate BP single-crystal flakes. We identify distinct peaks at 6 and 8.5 meV, which are attributed to the low-frequency interlayer phonon modes. The temperature-dependent red-shifts of both phonons below 150 K suggest the contribution of anharmonic phonon coupling. Following photoexcitation, free carriers undergo thermalization and relaxation processes through electron-phonon coupling, leading to transient red-shifts and sharp increases of the oscillator strengths of low-energy interlayer phonons. All these factors induce a significant enhancement of THz absorption in non-equilibrium. These findings underscore the significance of interlayer phonon dynamics in understanding many-body physics in BP.

## 2. Materials and methods

### A. Synthesis of BP flakes

BP single-crystal flakes were synthesized via chemical vapor transport (CVT) in a tube furnace with a single long heating zone. A mixture of red phosphorus, tin, and tin tetraiodide (acting as precursors and transport agents) was sealed in an evacuated quartz ampoule (~$10^{-2}$ Pa) and subjected to the following thermal profile: heating to 998 K (30 min hold), cooling to 600 K (120 min hold), and natural cooling to room temperature. The resulting BP flakes, with thicknesses of ~10 μm, were retrieved upon opening the ampoule.

### B. THz measurements

The BP sample was mounted freely on a holder inside an open-cycle cryostat. The pump beam was incident at an angle of nearly 30 degrees relative to the sample normal, with its polarization partially in- and out-of-plane. The probe THz beam was directed perpendicular to the sample surface (x-y plane) with a polarization approximately 60 degrees from the x-axis. To avoid water absorption, all measurements including THz generation, transmission and detection, were performed in a vacuum chamber maintained with a pressure of $10^{-7} \sim 10^{-6}$ mbar.

As for the THz setup, the fundamental laser source was a Ti:sapphire amplifier operating at a repetition rate of 1 KHz, with a pulse duration of 35 fs and a center wavelength of 800 nm. The output laser beam was split into three paths, serving as optical pump, THz generation and THz detection (sampling beam). THz probe pulses were generated via optical rectification in a 1 mm ZnTe (110) crystal. The detection of the THz electric field was achieved by electro-optic sampling (EOS) in another 1 mm thick ZnTe (110) crystal.

## 3. Results and Discussion

Figure 1(a) illustrates the experimental method for THz transmission measurements. We first carried out steady-state THz spectroscopy to explore the temperature-dependence of interlayer low-energy phonons below 10 meV. Photoexcitation of the system was achieved by time-resolved THz spectroscopy (optical-pump THz-probe, OPTP), with the pump wavelength at 800 nm (1.55 eV) and a fluence of 140 μJ/cm$^2$. Figure 1(b) depicts the THz-TDS results, including the reference (without sample) and the transmitted pulses through BP at 20 K. The steady-state THz spectra for both reference and transmitted pulses as a function of frequency/energy are obtained by fast Fourier transformation (FFT). Then the transmission can be calculated by dividing the FFT amplitude of the sample by that of the reference: $t(\omega) = I_{BP}(\omega)/I_{Ref.}(\omega)$. Figure 1(c) shows the transmission, revealing two dip features at ~6 and ~8.5 meV, corresponding to the absorption of two phonons.

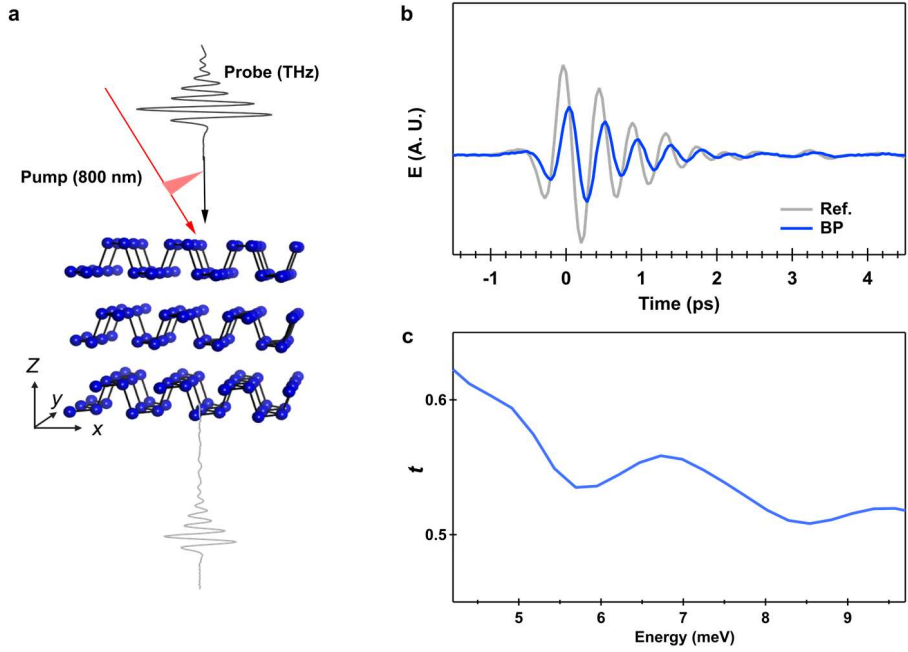

**Fig. 1.** THz spectroscopy on BP. (a) Sketch of time-resolved THz spectroscopy measurements. The transmitted THz probe is acquired to investigate the carrier and low energy phonon dynamics in BP. (b) Temporal evolution of THz electric fields (E) (time-domain spectroscopy, TDS) for reference (without the sample) and transmitted THz through BP measured at 20 K. (c) The transmitted THz spectrum of BP in frequency/energy domain by FFT analysis, $t(\omega) = I_{BP}(\omega)/I_{Ref.}(\omega)$, using the data in (b). Two transmission dips at ~6 and ~8.5 meV are identified.

The complex dielectric function, $\varepsilon(\omega)$, is derived from the complex transmittance, $T(\omega) = E_{BP}(\omega)/E_{Ref.}(\omega)$, which incorporates both the amplitude and phase changes obtained, following the relation [49]

$$T(\omega) = \frac{4\sqrt{\varepsilon(\omega)}}{\left(\sqrt{\varepsilon(\omega)}+1\right)^2} \exp\left[\frac{i\left(\sqrt{\varepsilon(\omega)}-1\right)d\omega}{c}\right] \qquad (1)$$

where $d$ is the sample thickness, and $c$ is the speed of light. Then the complex optical conductivity can be evaluated by the relation, $\sigma(\omega) = \sigma_1(\omega) + i\sigma_2(\omega) = i[(\varepsilon_\infty - \varepsilon(\omega)]\omega$. Eq. (1) describes the transmission of a single pass through the sample. In general, in optical measurements, multiple reflections will induce distortion in the wave profile due to interference effect, when the sample thickness is comparable to the wavelength of the light. Given that we observe well-defined THz TDS (Fig. 1b) and the sample thickness is nearly two orders of magnitude smaller than the THz wavelength, we reasonably assume that multiple reflections either play the same role as the original THz in the detection of low-energy excitations or are too weak to induce distortion; thus, we use Eq. (1) for data analysis.

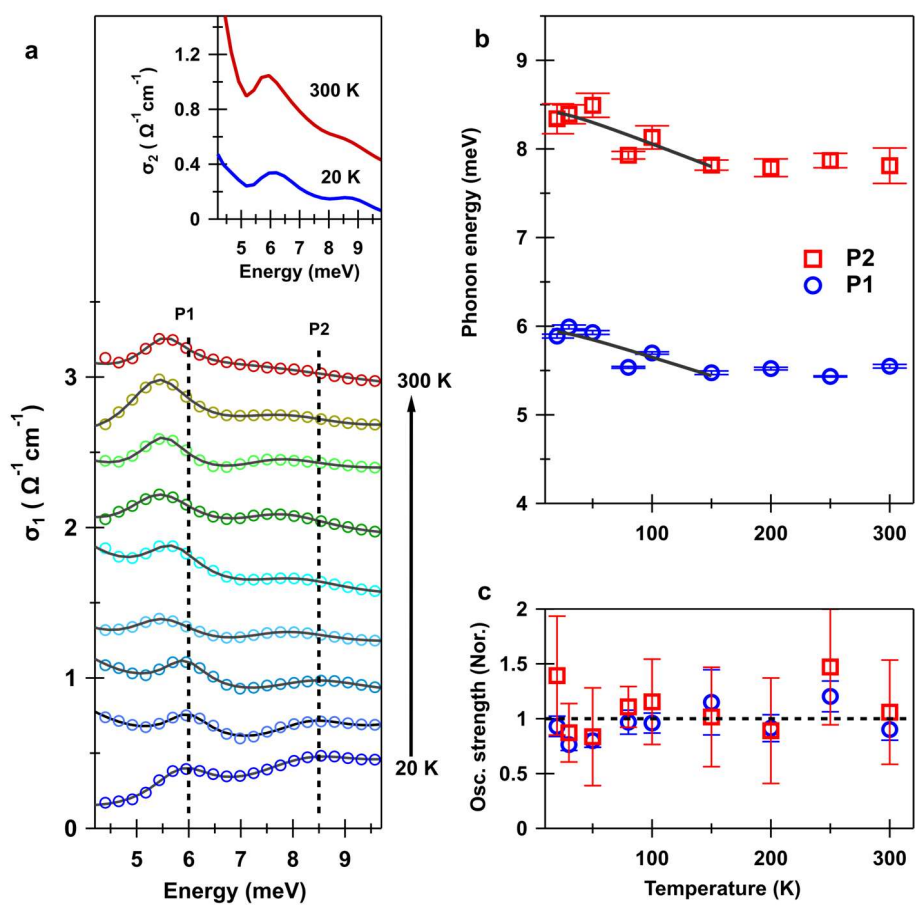

**Fig. 2.** Temperature-dependent THz phonons of BP. (a) Real part of the optical conductivity from 20 to 300 K. The dashed lines mark the two phonon positions at 6 (P1) and 8.5 meV (P2), respectively. The black curves are fit by the Drude-Lorentz model consisting of two phonon oscillators. The inset shows the representative imaginary part of the conductivity at 20 and 300 K, with inflection points around 6 and 8.5 meV. (b) Temperature evolution of P1 and P2 energies, extracted from Fig. 2a by the Drude-Lorentz fits. The black lines are fits of red-shifts below 150 K by the Klemens model. (c) Normalized oscillator strength as a function of temperature for P1 and P2 phonons.

Figure 2(a) displays the real part of the complex optical conductivity, $\sigma_1(\omega)$, at various temperatures. The observed peaks at 6 (P1) and 8.5 meV (P2) align with the calculated low-frequency interlayer infrared-active (IR) phonon modes, i.e., the in-plane shear mode (at 30 ~ 50 cm$^{-1}$, 4 ~ 6 meV) and the out-of-plane breathing mode (at 70 ~ 90 cm$^{-1}$, 8.5 ~ 11 meV), respectively [29,43,44]. The robustness of these modes is further confirmed by the imaginary part of the complex optical conductivity $\sigma_2(\omega)$, with inflection points at the 6 and 8.5 meV, as shown in the inset of Fig. 2(a). The experimental data are well described by the Drude-Lorentz model $\sigma(\omega) = \sigma_{dc}/(1-i\omega\tau_D) + \Sigma_m\{g_m\omega/[\omega\Gamma_m - i(\omega^2 - \omega_m^2)]\}$, where $\sigma_{dc}$ is the dc conductivity at zero frequency, $\tau_D$ is the scattering time, $g_m$ is the phonon oscillator strength, $\Gamma_m$ is the FWHM (full-width at half maximum) of the Lorentzian line, and $\omega_m$ is the phonon frequency/energy. Notably, it is evident that both P1 and P2 exhibit energy red-shifts and reduction of peak amplitude at higher temperatures. Figure 2(b) quantitatively plots the phonon peak energies as a function of temperature, obtained from the fits using the Drude-Lorentz mode. As temperature increases from 20 to 150 K, the peak energy of P1 decreases from 6 to 5.5 meV, and that of P2 decreases from 8.5 to 7.9 meV. For temperatures between 150 and 300 K, both P1 and P2 positions show very weak temperature dependence, with nearly no energy change in their peak values. The absence of robust changes of oscillator strengths of both P1 and P2 at different temperatures, as shown in Fig. 2(c), suggests that the phonon populations are not significantly affected by temperature. Here, the Drude component serves as a featureless background to extract the phonon properties. Our results are unable to explore the accurate free carrier response, given that the THz signal below 4 meV is limited, and other potential phonons below 4 meV cannot be excluded.

It has been well established that high-frequency intralayer phonons exhibit nonlinear temperature-dependent red-shifts, with only minor changes below 250 K, a behavior attributed to anharmonic phonon coupling [50,51]. In our results, the weak temperature dependence of P1 and P2 above 150 K, suggests a very weak anharmonic property for the low-frequency interlayer IR phonons in BP. This behavior is analogous to the low-frequency interlayer Raman breathing mode in BP, which also demonstrates weak anharmonic coupling above 150 K [44]. Notably, the red-shifts of P1 and P2 below 150 K provide direct evidence of the anharmonicity of the low-frequency interlayer phonon coupling, rather than the thermal expansion along the out-of-plane direction, which is weak and negligible [44]. The temperature dependence of the phonon anharmonicity based on a three-phonon process (one optical phonon decays into two acoustic phonons), can be described by the Klemens model [52]

$$\omega(T) = \omega_0 + A\left(\frac{2}{\exp(x)-1} + 1\right) \quad (2)$$

where A is an anharmonic constant, $x = \hbar\omega_0/2k_BT$, $\omega_0$ is the phonon frequency at 0 K. The fits for P1 and P2 red-shifts by the Klemens model are presented in Fig. 2(b). For simplicity, taking only the first term of the Taylor expansion, this model tends to a linear dependence with slopes of -5.4 and -3.8 μeV/K, for P1 and P2, respectively.

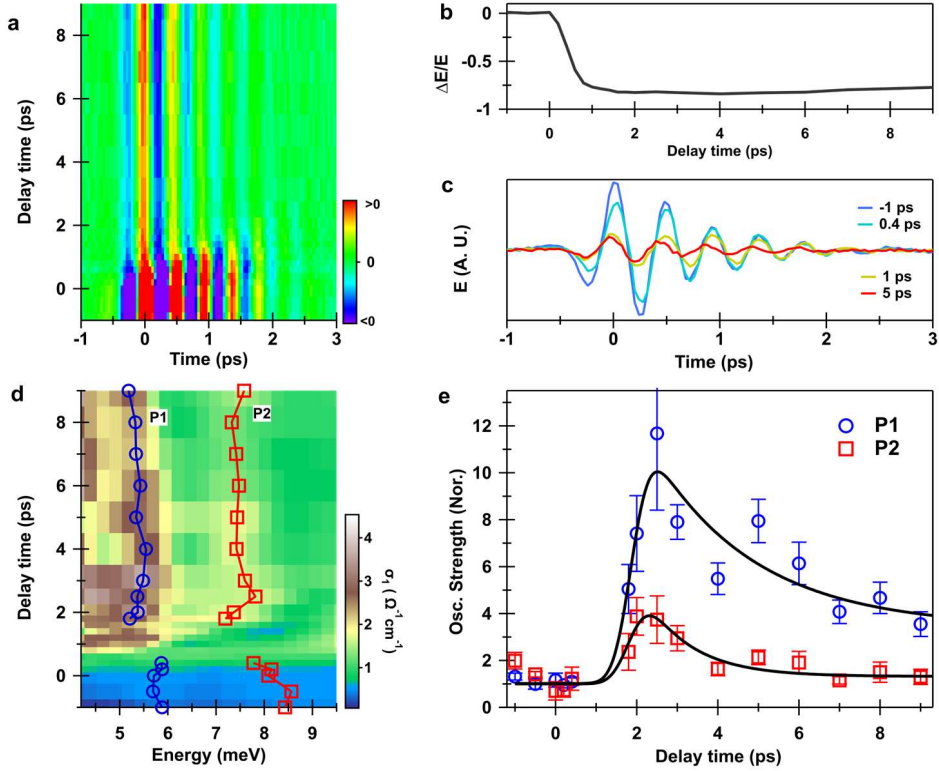

**Fig. 3.** Transient THz response of BP excited by pump pulses (800 nm, 1.55 eV), at a fluence of 140 μJ/cm². (a) Temporal evolution of THz-TDS by photoexcitation. (b) Changes of THz-TDS peak at 0 ps (ΔE/E) as a function of time, extracted from (a). (c) Typical THz-TDS curves at -1, 0.4, 1 and 5 ps. (d) 2D image of the real part of the transient optical conductivity. The blue and red curves are evolution of P1 and P2 phonon positions as a function of time, obtained from Drude-Lorentz fit. (e) Temporal evolution of phonon oscillator strength. The black curves are fits consisting of an error function (rising edge) and an exponential decay, with time constants of $\tau_{P1} = 2.6 \pm 1.5$ ps and $\tau_{P2} = 1.2 \pm 0.5$ ps.

Time-resolved THz spectroscopy is obtained by scanning the delay time between the optical pump and THz probe pulses. Figure 3(a) shows the THz-TDS as a function of pump-probe delay time at 20 K, revealing a sharp increase of the THz absorption after photoexcitation. Figure 3(b) plots a ΔE/E curve at the TDS peak position, indicating a transient decrease by 80% after 1.5 ps, compared to that of the unpumped THz signal at negative delays. This is further illustrated in Fig. 3(c), which compares typical THz-

TDS (horizontal cuts in Fig. 3(a)) at -1, 0.4, 1 and 5 ps. Figure 3(d) shows a 2D image of the evolution of the real part of the optical conductivity in frequency domain, calculated from Fig. 3(a). The dynamics from 0 to 1.5 ps are complex, primarily dominated by the extension of the Drude part and low-energy tail of high-frequency phonons, although a complete characterization of both components lies beyond the THz measurement regime in our experiments. Consequently, the Drude-Lorentz fits to extract photoexcited dynamics of P1 and P2 are performed after 1.5 ps. Notably, photoexcitation induces red-shifts in both P1 and P2 spectra, with their energies shifting from 6 to 5.5 meV (P1) and from 8.5 to 7.5 meV (P2), respectively. Additionally, the oscillator strengths of both phonon modes increase significantly, reaching nearly 10 (P1) and 4 (P2) times higher than their equilibrium values, with maximum enhancements occurring at 2.3 - 2.5 ps, as shown in Fig. 3(e). These indicate that photoexcitation effectively stimulates the anharmonic coupling and the population of low-frequency interlayer phonons.

Figure 4 establishes the ultrafast dynamics by extracting non-equilibrium kinetics at energies of 4, 5.5 (P1), 7.5 (P2) and 9.5 meV (low energy tail of high-frequency phonons), which contribute to the overall THz absorption. Although accurate free carrier response is unlikely obtained as mentioned above, the kinetics at 4 meV quantitively represents the trend of the Drude component, considering that its extension to a large THz region and the Drude background increases by nearly 5 times higher. Excited electrons are immediately pumped to higher-lying states by the pump pulses (1.55 eV), which significantly exceed the bandgap energy (~0.3 eV). The thermalization and relaxation processes to the minimum level, mediated by electron-phonon coupling, can be divided into three distinct regimes based on the experimental results. For simplicity, we focus on photoexcited electrons by assuming that excited holes undergo similar processes. The influence of radiative recombination in BP is neglected, as it has been found to occur on timescales on the order of 1 ns [53,54]. Electron thermalization can also be neglectable, as the timescale of electron-electron scattering is typically on the order of tens of fs [48,55,56]. In region (i), from 0 to 1.2 ps, photoexcitation immediately creates free electrons, leading to the extension of the Drude component. Electrons at high energy levels undergo thermalization process through the emission of high-frequency phonons. Thus, the transient THz spectra are predominantly influenced by the extension of Drude component and the low-energy tail of high-frequency phonons, with relatively weak contributions from P1 and P2. Toward the end of this regime (after 0.9 ps), the decay of high-frequency phonons starts emerging, resulting in the first peak feature in the P2 position. In region (ii), from 1.2 to 2.5 ps, where the electron energy lies between P1 and high-frequency phonons, the emission of P1 and P2 becomes dominant. This leads to the emergence of strong P1 and P2 peaks and their enhanced oscillator strengths in Fig. 3(e), accompanied by the continuous decay (absorption and dissipation) of high-frequency phonons. In region (iii), after 2.5 ps, when the electron energy drops lower than that of P1, phonon decays play a dominate role in the relaxation process. The decay time is inversely proportional to the phonon energy, resulting in a slow decay for P1 and a fast decay for P2. This relationship also explains the distinct decay times of the oscillator strengths in Fig. 3(e). More interestingly, the Drude component continues to increase in regions (i) and (ii) after photoexcitation, suggesting ongoing creation of free electrons. This behavior requires more precise investigations, particularly regarding the THz response below 4 meV.

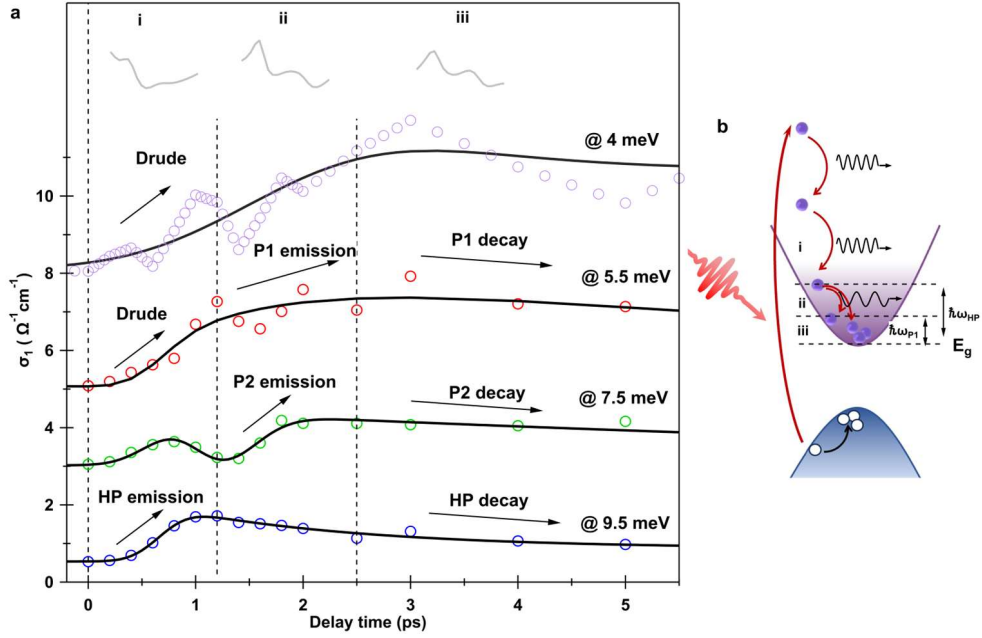

**Fig. 4.** Ultrafast carrier and THz phonon dynamics of BP after photoexcitation. (a) Cuts of the real part of the conductivity at 4 meV (trend of Drude component), P1 (5.5 meV), P2 (7.5 meV) and high energy (HP) tail (9.5 meV). The black curves are fits consisting of an error function (rising edge) and exponential decays. (i), (ii) and (iii) represent regions with different phonon components. (b) Optical pump (1.55 eV) generates over-gap excitation well above the bulk BP band gap (~0.4 eV). Excited electrons lying in the high energy conduction bands relax through the emission of phonons. Regions (i) and (ii) denote electrons with kinetic energy higher than high-frequency phonons (HP) and P1 ($\hbar\omega_{P1}$), respectively. Phonon emission dominate regions (i) and (ii), while phonon absorption and dissipation dominate region (iii) when electrons relax to energies below $\hbar\omega_{P1}$.

Time-resolved THz spectroscopy serves as a powerful tool for monitoring the ultrafast dynamics of photoexcited carriers and the transient evolution of low-frequency phonon modes. The observed transient red-shifts of low-frequency interlayer phonons on the timescale of few ps in BP are unlikely to be induced by photoinduced interlayer thermal expansion, as ultrafast electron diffraction (UED) measurements have demonstrated that such thermal expansion occur on a much longer timescale, extending up to tens of ps [57]. Considering that the strong interlayer coupling arises from the direct wavefunction overlap of lone electron-hole pairs, rather than the vdW interactions [5,42], transient modifications of the interlayer coupling are expected in the presence of photoexcited free carriers. UED has also revealed that photoexcitation-induced phonon population and electron-phonon scattering significantly drive the nonequilibrium lattice dynamics in BP [58]. Our results provide compelling evidence for a close relationship between photoexcited electrons and the dynamics of low-frequency interlayer phonons. Furthermore, the rapid population of high-frequency phonons immediately following the initial photoexcitation (region (i) in Fig. 4), possibly suggests an interlayer origin and an electron-coupled mechanism, given that above-gap excitation can induce an initial interlayer contraction on a similar timescale [57]. Further experimental and theoretical studies on electron-phonon coupling, particularly the relationship between interlayer phonon modes and transient properties, such as bandgap renormalization, carrier multiplication, and 3D screening [59,60,61], are still required to elucidate the many-body effects in BP.

## 4. Conclusions

In conclusion, we observe two low-frequency interlayer phonon modes at 6 and 8.5 meV in BP using steady-state THz spectroscopy. These modes exhibit temperature-dependent red-shifts below 150 K, which can be attributed to anharmonic coupling. Moreover, through time-resolved THz spectroscopy, we observe significant photoinduced THz absorption, enhancements in the phonon anharmonicity and the oscillator strength of low-energy interlayer phonons in non-equilibrium states. These effects arise from ultrafast photoexcited carrier thermalization and relaxation via electron-phonon coupling. Our findings provide deeper insights into the strong interlayer coupling in BP, which is crucial for advancing electronic and photonic applications.

**Funding.** National Natural Science Foundation of China (NSFC) (2250710126,